\documentclass[twocolumn,nofootinbib,superscriptaddress]{revtex4-1}

\pdfoutput=1

\usepackage{slashed}

\usepackage[normalem]{ulem}

\usepackage{color}

\usepackage{epsfig}
\usepackage{epstopdf}
\usepackage{epsf}
\input{epsf.sty}
\usepackage{graphicx,amsmath,amssymb}
\usepackage{epsfig}
\allowdisplaybreaks

\def\ei{\end{itemize}}
\def\be{\begin{equation}}
\def\ee{\end{equation}}
\newcommand{\bea}{\begin{eqnarray}}
\newcommand{\eea}{\end{eqnarray}}

\usepackage{bbm,bm,amsmath,amssymb}

\usepackage{hyperref}

\def\K{{K\"{a}hler}}

\newcommand{\rf}[1]{(\ref{#1})}
\usepackage{float}

\begin{document}

\title{4d models of dS uplift in KKLT}
\author{Renata Kallosh}
\email{kallosh@stanford.edu}
\affiliation{Stanford Institute for Theoretical Physics and Department of Physics, Stanford University, Stanford,
CA 94305, USA}
\author{Andrei Linde}
\email{alinde@stanford.edu}
\affiliation{Stanford Institute for Theoretical Physics and Department of Physics, Stanford University, Stanford,
CA 94305, USA}
\author{Evan McDonough}
\email{evan\_mcdonough@brown.edu}
\affiliation{Department of Physics, Brown University, Providence, RI, USA. 02903}
\author{Marco Scalisi}
\email{marco.scalisi@kuleuven.be}
\affiliation{Institute for Theoretical Physics, KU Leuven, Celestijnenlaan 200D, B-3001 Leuven, Belgium}

 \begin{abstract}

It was shown in \cite{Kallosh:2018wme} that the modified 4d version of the KKLT model proposed in \cite{Moritz:2017xto} is inconsistent for large values of the parameter $c$ advocated in \cite{Moritz:2017xto}, since there is a point in the moduli space where  $|D_SW|^2$ vanishes. The authors responded with yet another modification of the 4d KKLT model \cite{Moritz:2018ani}. However, for large $c$, this model suffers from an even worse problem: not only is there a point in the moduli space where $|D_SW|^2$ vanishes, there is also a region in the moduli space where $|D_SW|^2$ is negative.  Meanwhile for small $c$ these models have dS vacua.  We construct improved  models, which are fully consistent for all values of parameters, just as the original version of the KKLT model using a nilpotent superfield. These models have a family of dS vacua for a broad range of parameter values.   Thus,  the results of the  analysis of all presently available  consistent generalizations of the 4d KKLT model, in the domain of their validity,  confirm the existence of dS vacua in the KKLT scenario. 
 \end{abstract}

\maketitle


\section{Introduction}

The recent KKLT debate between   \cite{Kallosh:2018wme} and \cite{Moritz:2017xto}   is currently  involving two sides of the story. One is based on the 10d analysis presented in \cite{Moritz:2017xto}, recently summarized  in \cite{Moritz:2018ani}. In the latter paper,  the authors  reviewed their previous long paper \cite{Moritz:2017xto}, presenting a short version of their arguments in 10d. In our opinion, expressed in \cite{Kallosh:2018wme}, their results are based on several unjustified and debatable assumptions. In the absence of actual computations in \cite{Moritz:2017xto}, the status of the 10d arguments  will remain inconclusive until such an explicit analysis is actually performed. A similar conclusion was reached in \cite{Cicoli:2018kdo}.

Meanwhile, in 4d the situation is more transparent, being based on  4d supergravity with a nilpotent multiplet $S$, representing an anti-D3 brane. Here the explicit equations can be easily checked. In  \cite{Moritz:2017xto} a \K\,  potential $K$ and a superpotential $W$ were given which were supposed to confirm the 10d analysis in \cite{Moritz:2017xto}. The authors of  \cite{Moritz:2017xto} conceded in \cite{Moritz:2018ani} that  their first model (we will call it v1) is inconsistent for the values of the parameters $|c A | \gg b$ advocated in  \cite{Moritz:2017xto}.

Therefore they have now proposed another version of their model in \cite{Moritz:2018ani} (we will  call it v2). We will study the model v2 and perform the corresponding analysis of $|D_SW|^2$ as a function of $T$. Surprisingly, we see again that $|D_SW|^2$ can vanish and even be negative, which invalidates the new model proposed in \cite{Moritz:2018ani}   for large $c$. Meanwhile for small $c$ this model has a family of $dS$ vacua.  This is in  contradiction with the claim in \cite{Moritz:2018ani} that their model v2 is `better' than their model v1, and `can match the ten dimensional result'.   On the contrary, we find that the second model, in the domain of its validity, supports the standard conclusion of the existence of dS vacua in the KKLT model. 

Finally, we also suggest how to modify the models in \cite{Moritz:2017xto,Moritz:2018ani} so that the positivity of $|D_SW|^2$ is preserved  for all values of their parameters.  In these new models we still find dS solutions. This means that even when a model consistently deviates from the original KKLT scenario, metastable dS vacua are still preserved.

\section{Model v1 and model v2}
\subsection{Model v1}

The original version of the KKLT scenario in the formulation where the anti-D3 brane is represented via a nilpotent multiplet $S$ is given by \cite{Ferrara:2014kva,Kallosh:2014wsa,Bergshoeff:2015jxa}
\be\label{WKKLT}
W= W_0 +Ae^{-aT} +  bS ,
\ee
and the \K\, potential which can be either 
\be K=-3\log\left(T+\bar T \right) +{S \bar S}, 
\ee
or
\be
K=-3\log\left(T+\bar T -{S \bar S}\right) \ .
\ee
The modification proposed in \cite{Moritz:2017xto}
introduces an extra term $ c Ae^{-aT} S$ in the superpotential, with an extra parameter $c$, which is supposed to describe effects of backreaction\, 
\be\label{W1}
W= W_0 +Ae^{-aT} + c Ae^{-aT} S+ bS \ .
\ee
It was argued in \cite{Moritz:2017xto} that $|c A | \gg b$.  This argument, which is also central to their 10d approach, does not seem well motivated, because it would imply that the backreaction to the anti-D3 brane is much greater than the main effect of the anti-D3 brane   \cite{Kallosh:2018wme, Cicoli:2018kdo}. We studied the general case, including $|c A | \ll b$ as well as $|c A | \gg b$.

After \K\, transformation 
\be
(c Ae^{-aT} +b) S \rightarrow \tilde S \ ,
\ee
an equivalent model is (ignoring   tilde)
\be\label{wnob}
W= W_0 +Ae^{-aT} + S ,
\ee
\be\label{wv1}
K=-3\log\left(T+\bar T -{S {\bar S}
\over  |c Ae^{-aT} +b|^2}\right) .
\ee
Note that the denominator in \rf{wv1} is a perfect square, $|c Ae^{-aT} +b|^2 \equiv (c Ae^{-aT} +b) (c Ae^{-a\bar T} +b)$. It is positive everywhere except the point
\be
\label{sing}
T_0= \frac{1}{a}\ln\left(-\frac{cA}{b}\right)\, ,
\ee
where it vanishes. This makes the use of the nilpotent multiplet $S$ in the  model \rf{W1} inconsistent \cite{Kallosh:2018wme}.    This violation of the consistency requirement occurs at large volume, e.g.~$|T_0| \simeq 120$ for $\{a, A,b,c \} = \{0.1, 1,10^{-5},1 \} $, and hence is a problem in precisely the region of moduli space where the nilpotent multiplet is expected to provide a valid effective field theory description of an anti-D3 brane.

If one disregards this problem of the model  proposed in  \cite{Moritz:2017xto} and calculates the resulting potential, one finds that the theory does contain a large family of dS vacua, some of which have not been found in the previous works \cite{Kallosh:2018wme}. The authors of  \cite{Moritz:2017xto} argued that we found dS states for $c = 1$, whereas we found dS states for a very broad set of parameters, starting from very small $c$, all the way to $c = 10^{4} \gg b$.
\vspace{-0.7cm}
\subsection{Model v2}
Consider a new model v2  \cite{Moritz:2017xto}. We are given the superpotential \rf{wnob}, and \K\, potential
\be
K=-3\log\left(T+\bar T -{S \bar S
\over \tilde b^2}\right) ,
\ee 
with $T = t+i\theta$ and $\tilde b^2$  depending on some arbitrary functions $f(T+\bar T)$ and $g(T+\bar T)$ such as
\begin{equation}
\label{btilde}
\tilde b^2=b^2 + b(f(T+\bar T) e^{-aT}+c.c.) + g(T+\bar T)e^{-a(T+\bar T)}\,.
\end{equation}
However, the model called `better' has a particular choice of these two functions,
$f(T+\bar T)=\bar c$ and \mbox{$g(T+\bar T)=g_1 \cdot (T+\bar T)$}, where $g_1$ is a constant.
So we proceed from there and define the v2 \K\, as follows. First, we reorganize the expression to take out the nilpotent field from the $\log$
\be
K=-3\log\left( (T+\bar T )\left(1 -{S \bar S
\over (T+\bar T)\tilde b^2}\right)\right) ,
\ee
so that
\be
K=-3\log (T+\bar T ) - 3 \log \left (1 -{S \bar S
\over (T+\bar T)\tilde b^2}\right)\,,
\ee
and, finally 
\be
K=-3\log( T+\bar T ) +  {3\over (T+\bar T)\tilde b^2} S \bar S .
\ee
Thus we have 
\be
K^{S\bar S}= {T+\bar T\over 3} \bigl(b^2 + b(\bar c e^{-aT}+c.c.) + g_1 (T+\bar T)e^{-a(T+\bar T)}\bigr)  .
\ee

Thus, in order to disprove KKLT, the authors of  \cite{Moritz:2017xto} introduce a highly sophisticated modification of the original KKLT model  \cite{Ferrara:2014kva,Kallosh:2014wsa,Bergshoeff:2015jxa}, containing not one, but two extra parameters, $c$, and $g_{1}$. 

We are now ready to compute the supersymmetry breaking in the $S$ direction, that is \mbox{$|D_S W|^2\equiv D_S W K^{S\bar S} \bar D_{\bar S} \bar  W$}, and to study it at $S=0$. Note that the sign of $|D_S W|^2$ is determined by the sign of $K^{S\bar S}$. Since $D_S W = 1$, we find 
\be
|D_S W|^2=  {T+\bar T\over 3}\bigl(b^2 + b(\bar c e^{-aT}+c.c) +  2 g_1 t e^{-a(T+\bar T)}\bigr).
\ee
In terms of $t, \theta$ we can present it as follows at $t>0$ and in the simple case $c=\bar c$
\be\label{CCCC}
|D_S W|^2=  {2t \over 3}\bigl(b^2 +  2 b c  e^{-at } \cos a \theta + 2 g_1  t e^{-2a t}\bigr)\,.
\ee
If we take $\theta =\pi/a$, we find
\be
|D_S W|^2=  {2t \over 3} \bigl(b^2 -  2 b c  e^{-at }  + 2 g_1 t e^{-2a t}\bigr)\,.
\label{square}\ee
As before, for positive $bc$ we find that the vanishing of $|D_S W|^2$ can be achieved in the complex plane of $T$ with positive $T+\bar T$. This makes the model v2 inconsistent to the same degree as the model v1, where the corresponding equation was
\be
|D_S W|^2= {2t\over 3} |b + c  e^{-aT } |^2   > 0 .
\ee
Thus, if we take $\theta =\pi/a$ in \rf{CCCC}, we find a point in field space where $|D_S W|^2=0$, which invalidates the model  for $c>{b\over 2}$.
A similar conclusion is reached for $|c|$ in the general case $c =|c| e^{i\gamma}$, but for a different value of $\theta$, depending on the phase  $\gamma$.

In addition to a possibility that in model v2 \mbox{$|D_S W|^2=0$} is possible, one finds that at small $t$ where the third term in eq. \rf{square} is small, the expression in \rf{square}  is negative, and the model is inconsistent for $c>{b\over 2}$.   Meanwhile for $c < b/2$  the corrections proportional to $c e^{-aT} $ are exponentially suppressed as compared to $b$. Therefore the only relevant ``backreaction'' term is the one proportional to $g_{1}$. We checked that dS vacua exist in this model even if  $g_{1}\gg b$, just as in the previous model  \rf{W1}.

\vspace{-0.4cm}
\section{Consistent generalizations of the KKLT model}
As we have shown above, both versions of the modified KKLT construction proposed in 
\cite{Moritz:2017xto, Moritz:2018ani} are inconsistent  for some values of their parameters,  because both of them violate the consistency requirement for the description of the anti-D3 branes in terms of the nilpotent multiplet. Now we will solve this problem  and propose some models which are consistent for all values of their parameters.  

We will keep the original version of $W$ \rf{wnob} and make a minor modification of the \K\ potential
\be\label{wv3}
K=-3\log\left(T+\bar T -{S {\bar S}
\over  |c Ae^{-aT} +b|^2+\beta \,c^2 A^2  e^{-a(T+\bar T)}}\right) \,
\ee
where $\beta$ is some positive number. This immediately makes $K^{S\bar S}$ strictly positive definite, which avoids all inconsistencies of the models of  \cite{Moritz:2017xto,Moritz:2018ani} for any choice of $\beta > 0$. This model falls in the category of models previously studied in \cite{McDonough:2016der,Kallosh:2017wnt}.

Note that for $0<\beta\ll 1$, the KKLT potential in this model practically coincides with the potential obtained in our paper  \cite{Kallosh:2018wme}. Thus all our previous results about the existence of dS vacua contained in  \cite{Kallosh:2018wme} are confirmed for a large range of parameters, without any problems with the nilpotent multiplet $S$ encountered in  \cite{Moritz:2017xto,Moritz:2018ani}.

Yet another, even simpler model, is described by
 $W$ \rf{wnob} and \K\ potential 
\be\label{wv4}
K=-3\log\left(T+\bar T -{S {\bar S}
\over  |b|^2+ |c|^2 e^{-a(T+\bar T)}}\right), \,
\ee
which amounts to the choice $f=0$ and $g=|c|^2$ in \eqref{btilde}. $K^{S\bar S}$ is strictly positive definite, which makes it consistent for any choice of $\{b,c\}$. This model also belongs to the class of models previously studied in \cite{McDonough:2016der,Kallosh:2017wnt}.

\begin{figure}[h!]
\vspace*{3mm}
\begin{center}
\includegraphics[width=8cm]{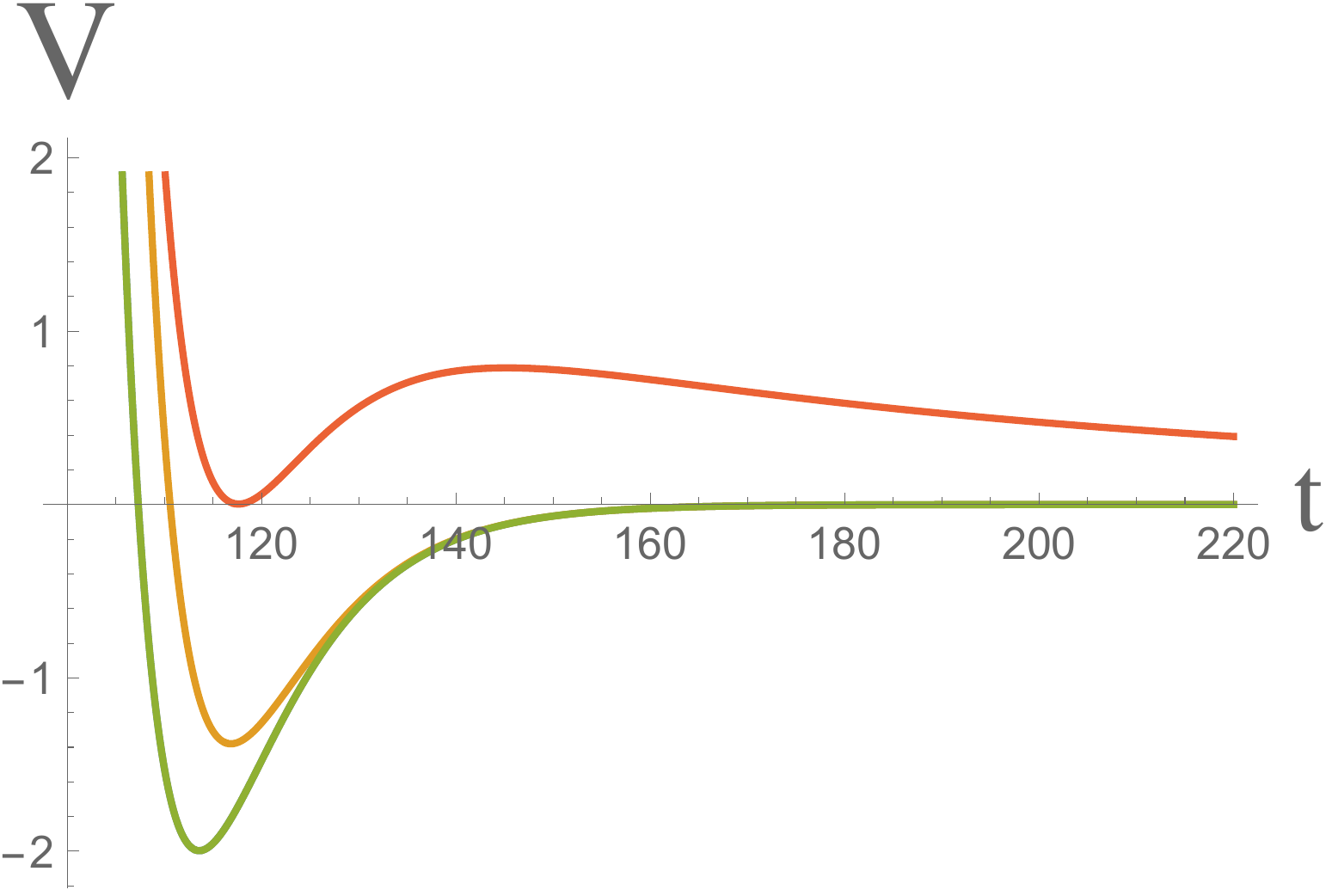}
\caption{The potential for the model \rf{wv4} (multiplied by $10^{15}$) for  $A=1$, $a=0.1$, $W_0=-10^{-4}$, $c= 1$. The green (lower) line shows the potential with a supersymmetric AdS minimum prior to uplifting, at $b = c = 0$. The second (yellow),  line shows the potential at $b = 0$ uplifted by increase of $c$ to  $c = 1$. This does not uplift the potential to dS. Finally, the upper (red) line shows the potential with a dS (nearly Minkowski) minimum for $c = 1$, $b =1.51\times 10^{{-5}}$.}
\label{f2}
\end{center}
\vspace{0cm}
\end{figure}

This model, just as the original KKLT model where the anti-D3 brane is represented via a nilpotent multiplet $S$  \cite{Ferrara:2014kva,Kallosh:2014wsa,Bergshoeff:2015jxa}, has dS vacua for a broad choice of its parameters.
We illustrate our results for the case $c=1$, $b=1.51 \times 10^{-5}$ in Fig. \ref{f2}. It should be compared to Fig. 2 in our previous paper  \cite{Kallosh:2018wme}.

\section{Conclusion}
\label{sec:conclusion}

In this note we have revisited the inconsistencies of the 4d supergravity analysis done in \cite{Moritz:2017xto}, found previously in \cite{Kallosh:2018wme} for large values of $c$ advocated in \cite{Moritz:2017xto}. The authors of \cite{Moritz:2017xto} responded to  \cite{Kallosh:2018wme} with a revised model \cite{Moritz:2018ani}, which we have referred to as model v2. In this note we have found the same, and yet further, inconsistencies in the revised model at $c >b/2$: (1) as in the original model the supersymmetry breaking $|D_S W|^2 = K^{S \overline{S}}$ vanishes at a point in moduli space, and (2) $K^{S \overline{S}}$ can in fact become \emph{negative}. Such models do not have an embedding in de Sitter supergravity, at least as it currently formulated, as is the case for the model given in \cite{Moritz:2017xto}.

 However, with simple modifications, which we outline in section III, these inconsistencies can be removed, leading to a family of dS solutions without problems (1) or (2) mentioned above.  The results of the detailed analysis of various consistent generalizations of the 4d KKLT models, in the domain of their validity, invariably confirm the existence of dS vacua in the KKLT scenario.

\acknowledgments

We would like to thank Shamit Kachru,  Ander Retolaza, Eva Silverstein, Sandip Trivedi, Thomas Van Riet, and Timm Wrase for helpful comments and discussions. The work  of RK and AL is supported by SITP,  by the NSF Grant PHY-1720397, and by the Simons Foundation grant.  EM is supported in part by the National Science and Engineering Research Council of Canada via a PDF fellowship.  MS is supported by the Research Foundation - Flanders (FWO) and the European Union's Horizon 2020 research and innovation programme under the Marie Sk{\l}odowska-Curie grant agreement No. 665501.
%

\vspace{0.5cm}

\bibliography{lindekalloshrefs}
\bibliographystyle{utphys}

\end{document}